%% file: EBsaliency.tex
\documentclass[journal]{IEEEtran} 
\usepackage{cite}
\usepackage[help]{epspdfconversion}
\usepackage{graphicx}
\ifCLASSOPTIONcompsoc
\usepackage[caption=false,font=normalsize,labelfont=sf,textfont=sf]{subfig}
\else
\usepackage[caption=false,font=footnotesize]{subfig}
\fi

\usepackage{booktabs,caption,threeparttable}
\usepackage{multirow}   
\usepackage{bm}
\usepackage[cmex10]{amsmath}
\usepackage{array}
\usepackage{algorithm}
\usepackage{algpseudocode}
\usepackage{xspace}
\usepackage{siunitx}
\usepackage{booktabs}
\usepackage{mathtools}
\usepackage{soul}
\usepackage{url}
\usepackage{hyperref}
 \hypersetup{
     colorlinks=true,
     linkcolor=blue,
     filecolor=blue,
     citecolor=blue,      
     urlcolor=magenta,
     }
\usepackage{xcolor}
\usepackage{times}
\usepackage{bbold}
\usepackage{float}

\input{abbrevs}

\definecolor{redsol}{RGB}{220,50, 47}
\definecolor{violetsol}{RGB}{108, 113, 196}
\definecolor{bluesol}{RGB}{38 139 210}
\definecolor{cyansol}{RGB}{42 161 152}

\algnewcommand{\aAnd}{\textbf{and}\xspace}
\algnewcommand{\aOr}{\textbf{or}\xspace}

\DeclarePairedDelimiter\abs{\lvert}{\rvert}%
\DeclarePairedDelimiter\norm{\lVert}{\rVert}%
\makeatletter
\let\oldabs\abs
\def\abs{\@ifstar{\oldabs}{\oldabs*}}

\let\oldnorm\norm
\def\norm{\@ifstar{\oldnorm}{\oldnorm*}}
\makeatother

\begin{document}
\title{An event-based implementation of saliency-based visual attention for rapid scene analysis}
\author{
Camille Simon Chane$^{1}$, 
Ernst Niebur$^{2}$,
Ryad Benosman$^{3}$,
Sio-Hoi Ieng$^{3}$\\
$^{1}$ETIS UMR 8051, CY Cergy Paris University, ENSEA, CNRS\\ 
$^{2}$Dept. of Electrical and Computer Engineering and Dept. of Neuroscience, Johns Hopkins University\\
$^{3}$Dept. of electrical engineering, Sorbonne University
} 
\maketitle
 
 \begin{abstract}
Selective attention is an essential mechanism to filter sensory input and to select only its most important components, allowing the capacity-limited cognitive structures of the brain to process them in detail. The saliency map model, originally developed to understand the process of selective attention in the primate visual system, has also been extensively used in computer vision. Due to the wide-spread use of frame-based video, this is how dynamic input from non-stationary scenes is commonly implemented in saliency maps.
However, the temporal structure of this input modality is very different from that of the primate visual system. Retinal input to the brain is massively parallel, local rather than frame-based, asynchronous rather than synchronous, and transmitted in the form of discrete events, neuronal action potentials  (spikes). These features are captured by event-based cameras. We show that a computational saliency model can be obtained organically from such vision sensors, at minimal computational cost. We assess the performance of the model by comparing its predictions with the distribution of overt attention (fixations) of human observers, and we make available an event-based dataset that can be used as ground truth for future studies.

\comment{
To reproduce this mechanism on machine, the concept of saliency maps has been designed in computer vision to quantify and
predict where in an image relevant information are captured. Extensive works have been produced over the past decades however with little breakthrough when it comes
to extent the concept of salient information to dynamic context of video sequences. This work focuses on the use of an event-based vision sensor that natively implements    
the magno-cellular pathway in the visual cortex that captures transient information. We show that a spatial and time continuous saliency model can be easily and almost
for free built from such vision sensor. We also evaluated the model by providing large event-based datasets that we have acquired, where ground truth has been built by
tracking eye fixations from a cohort of 20 subjects. 
}
\end{abstract}

\section{Introduction}
The primate visual system is capable of dealing with extremely large amounts of data coming from the eyes without being overwhelmed. Visual information is pre-processed in the retina and then sent to the brain {\em via} the optic nerves. A simple calculation shows that, seen as an information transmission channel in the Shannon sense, each optic nerve has a capacity of $\approx 10^8$ bits per second~\cite{Koch2004,LeCallet_Niebur13}. The channel is discretized since information is coded in terms of the presence and absence of discrete events, the action potentials of retinal ganglion cells. The brain deals with this deluge of information by the mechanisms of visual selective attention that allocate resources to process preferentially information deemed relevant, and suppresses or discards the remainder.

The concept of visual selective attention also plays a critical role in computer vision, machine learning, and robotic research, which face the same problem of overwhelming amounts of data arriving from high-throughput vision sensors. 
The ability to process all these data at the rate of their acquisition is particularly critical for autonomous systems that are subject to tight power and computational constraints. How a human brain is capable to process all of the incoming information and achieve complex tasks by consuming as little as 20W is not understood, but selective attention is likely one of the reasons: early selection of relevant information and discarding all other presumably saves costly computational resources. 

Selective attention is a complex process that plays a crucial role at many levels of perception and cognition. It is useful to distinguish between top-down and bottom-up attention. The latter can be described as data-driven: where bottom-up attention is directed is determined by the visual input alone. In contrast, top-down attention also depends on the internal states of the observer, for instance their (or, in the case of machine vision, its) goals. While progress is made in understanding mechanisms of top-down attention~\cite{Baluch_Itti11,Connor_etal04,Mihalas_etal11b,Usher_Niebur96b}, it is more difficult to study because it is usually much easier to control visual input than the internal states of an observer. This is one of the reasons why bottom-up attention has been studied in much more detail and why it is also the focus of the present study. 

Highly influential conceptual models of visual selective attention were developed in the 1980s which introduced Feature Integration Theory~\cite{Treisman1980} and the concept of the saliency map~\cite{Koch1985}. The latter was moved from the stage of a conceptual idea to a quantitative, testable computational model~\cite{Niebur_Koch96b,Itti1998,Itti_Koch01}; review:~\cite{Niebur07c}, which is part of the pedigree of the present study. 

These early models, as well as many contemporary versions, were designed for static visual scenes (although an early version of a dynamic saliency map was already introduced in 1996~\cite{Niebur_Koch96b}). In the real world, however, agents are interacting with a continuously changing environment, making the integration of temporal information in the computation of saliency a necessity. It is therefore not surprising that saliency map models have incorporated the effects of visual changes on the control of selective attention, e.g.~\cite{Itti2004,Itti05,Rosenholtz1999,Molin_etal21}. The vast majority of such models are frame-based, being applied to the standard representation of dynamic scenes (e.g. in video) in terms of series of image "frames" shown at a sufficiently high rate to result in the perception of smooth movement. 
Of course, primate biology has no concept of image frames. Instead, sensory input is dynamic and sent asynchronously from different retinal ganglion cells to the brain.

In the present study, visual input is not represented in the form of standard frame-based video cameras but, instead, generated by event-based vision sensors such as the Dynamic Vision Sensor~\cite{Lichtsteiner2008}. Event-based vision sensors asynchronously capture spatio-temporal constrast changes, akin to the magnocellular pathway in primate vision. They are sensitive to changes due to motion, onsets and offsets. 
Visual information going beyond the detection of changes, providing information about the local intensity at each pixel is needed for tasks like object recognition and many others. In the primate visual system, this is mainly represented in the parvocellular pathway. We use a variant of event-based cameras, the Asynchronous Time-based Image Sensor (ATIS)~\cite{Posch2011} that represents not only the time when a change is detected (loosely related to the magnocellular pathway as discussed) but, using a pulse-width modulation code, also the pixel value at the time of each change, loosely related to the parvocellular pathway. These signals are, again asynchronous and binary, akin to action potentials generated in retinal ganglion cells.  In the current study, a component of the parvocellular pathway, color sensitivity, is missing, but newer versions of the ATIS sensor do provide this feature~\cite{Marcireau_etal18}, and future work can incorporate it into saliency computations. 

Of particular interest for the present study is that event-based vision sensors natively capture fast changes in the visual input which, we argue, is a highly efficient way to compute saliency. While temporal change is only one of several contributions to bottom-up saliency~\cite{Niebur_Koch96b,Molin_etal15a}, it is a highly important one~\cite{Itti05,Rosenholtz1999} even though trained artificial neural networks seem to weigh dynamical changes less than static features~\cite{Tangemann_etal20}.

One main contribution of this work is the creation of an event-based pipeline used to compute a saliency score for each event, at the same rate as events occur. This makes the event-based sensor a native dynamical visual saliency identifier.
An event's saliency score is the response of spatiotemporal filters that average the number of events that occurred within a spatiotemporal neighborhood of the current event. The model has a bottom-up (feed-forward) architecture and, importantly,  it does not require any training. Since event-based encoding of information incurs only minimal redundancy, we propose an easy to implement and computationally highly efficient event-based saliency map on a generic consumer-grade computer.

Our second main contribution is the compilation of a dataset of event-based camera data obtained by recording dynamic visual scenes. They are combined with eye-movement data collected from human observers while they free-view the corresponding visual scenes. We use these data to evaluate the fixation-prediction performance of our model, as well as that of other dynamical saliency models. We show that the proposed model outperforms state-of-the-art saliency models when applied to spatiotemporal or dynamic contents. Since, to our knowledge, currently no such dataset has been published, we make our dataset freely available. 

\section{Related Work}

While most early saliency map models were designed for still images~\cite{Itti1998,Itti_Koch01}, the importance for temporal change for attracting attention  led to the development of computational models that use spatio-temporal filters~\cite{Parkhurst02,Gao2008} or localized temporal change ("flicker")~\cite{Niebur_Koch96b,Itti05} as components contributing to local saliency. A related approach is the use of statistical analyses to introduce dynamical components in saliency models. Bayesian inference is the root of the method proposed in~\cite{Itti2005}, 
where the concept of Bayesian surprise is used to characterize spatio-temporal change. Motion features can then be captured as  elements of dynamic saliency. Another related approach~\cite{Rosenholtz1999} relies on saliency generated from outliers in a pre-established/learned distribution of motion features.

Feature Integration Theory~\cite{Treisman1980} has spawned a large set of saliency models that can be classified as "feature-based". The basic elements in these models are elementary visual features, \eg oriented line segments, colored areas \etc It is known, however, that primates use more complex concepts to structure their visual input to increase efficiency of scene understanding. Perceptual organization based on Gestalt psychology~\cite{Rubin21} is the basis for models implementing these ideas. An important concept is that of proto-objects, the non-semantic precursors of object representations~\cite{Rensink00a}. (Proto-)Object-based scene organization has been exploited for a biologically plausible  saliency model and shown to exceed performance of purely feature-based models in the prediction of overt attention (fixations)~\cite{Russell2014}. While the original model~\cite{Russell2014} did not include temporal information, a generalization~\cite{Molin_etal13,Molin_etal15a} implements the use of spatio-temporal filters over a subset of consecutive frames in a sliding window. Dynamic information and the concept of motion then becomes part of the saliency computation.  The method is strictly feed-forward and does not require any training which makes it possible to implement it on dedicated hardware (FPGA) that can run as a standalone real-time device~\cite{Molin_etal21}.

Most of these methods are built on biological evidence and are neuromorphic in this perspective; however, they are still frame-based rather than acquired in an event-based, asynchronous way.
Computation of the saliency model requires substantial pre-processing due to the redundant information contained in frames. An alternative comes from event-based vision sensors that are now experiencing a growing acceptance in visual sensing. As we show in this work, they are new tools for tackling the problem of implementing visual attention in machines. As event-based vision sensors are designed to be driven by spatio-temporal contrast changes, roboticists~\cite{Iacono2019,Gabbott_etal87} have developed event-based forms of proto-object based visual saliency.  

\section{Model}
\subsection{Event-based sensors}
Event-based sensors available on the market are mainly variations of the Dynamic Vision Sensor~\cite{Lichtsteiner2008}, with steadily improving spatial resolution in newer generations.
The core principle of these sensors is to detect intensity changes at the level of each individual pixel. A supra-threshold change from low-to-high intensity generates an ON event, and from high-to-low an OFF event. Since information is only transmitted when a temporal change occurs at any given pixel, both spatial and temporal redundancy are largely reduced in the acquired visual data. 
For this reason, the currently technologically available bandwidth allows update rates in the microsecond range. 

In addition to implementing this change detection algorithm, the ATIS camera also captures brightness (luminance) information~\cite{Posch2011}.  Each pixel of this sensor has two detectors: The first is the contrast change detector that produces pixel-wise ON/OFF events as described above. In addition, each event of this detector triggers a second detector which represents the luminance of the changed pixel by generating a pair of events whose time difference encodes the pixel luminance. The functionality of this mechanism is roughly analogous to the intensity component of the parvocellular pathway in the mammalian visual system, although the implementation (temporal difference coding \vs labeled line coding) is very different.

\subsection{Saliency model}
\label{sec:model}
\subsubsection{Basic saliency score}
We postulate that an event-based camera implementing temporal contrast change detection is a native dynamic visual saliency sensor. By collecting events in a two-dimensional spatial map and updating old values as new events come in, we are generating a spatiotemporal saliency map. A limitation is that with the currently used ATIS sensor only one submodality (intensity) contributes while traditional models combine multiple submodalities~\cite{Koch_Ullman85,Itti1998,Itti_Koch01}. Although this is an important constraint, it is not of a fundamental nature; it can be addressed by using other hardware sensors, for instance those including color information~\cite{Marcireau_etal18, Berner_Delbruck11}.

Let us assume that $e=\{x,y,t\}$ is the current event, occurring at time $t$ and position $(x,y)$. We define a spatiotemporal history of the event by the spatial neighborhood of size $r_v$ around $(x,y)$ and the duration $t_u$ before $t$. The saliency score at that time and position is the response of a filter that sums the number of events within the thus-defined spatiotemporal history  $(r_v,t_u)$ as:

\begin{equation}
S_{u,v}(x,y,t)=\sum_{i} \frac{\mathbb{1}_\sigma(x_i,y_i,t_i)}{(1+2r_v)^2},
\label{eq:saliency}
\end{equation}
where
\begin{equation}
\sigma =\{e_i \big| \ |x-x_i|+|y-y_i| \le r_v\textrm{ and } t-t_i \le t_u\}
\label{eq:set}
\end{equation}
and
\begin{equation}
\mathbb{1}_\sigma(x_i,y_i,t_i) =\left\{ \begin{array}{ll} 
1 \textrm{ if } (x_i,y_i,t_i) \in \sigma \\
0 \textrm{ otherwise}
\end{array}\right., 
\end{equation}
is the indicator function of the set $\sigma$. It is then "normalized" by the area of the spatial window of width  $(1+2r_v)$. We use the tarsier framework~\cite{Marcireau2020} without a decay mechanism to store the events in a 2D buffer where only the most recent event at each pixel is stored. This buffer has the same purpose as the one used in storing a "time-surface" introduced in~\cite{Lagorce2017}.

\subsubsection{Saliency at multiple scales}
Equation~\ref{eq:saliency} defines the saliency score computed for one spatial scale $r_v$ and one temporal window $t_u$. To capture changes at multiple spatiotemporal scales, we include six octaves in both space and time. Formally, we define
\begin{eqnarray}
r_v \in \{2^v\}_{0\le v \le 5} \\
t_u \in \{10 \times 2^u\}_{0\le u \le 5} \, \, {\rm ms}
\label{eq:scales}
\end{eqnarray}
and use $S_{u,v}$ from eq~\ref{eq:saliency} to define the spatiotemporal saliency $S_{ST}$ at position $(x,y)$ and time $t$ as,
\begin{equation}
S_{ST}(x,y,t)=\frac{1}{ST_{max}}\sum_{u=0}^5 \sum_{v=0}^5 S_{u,v}(x,y,t),
\label{eq:SST}
\end{equation}
where the division by $ST_{max}=\max_{x,y,t}(S_{ST})$ normalizes $S_{ST}$ to the range $[0,1]$.

\begin{figure}
\includegraphics[width=\columnwidth]{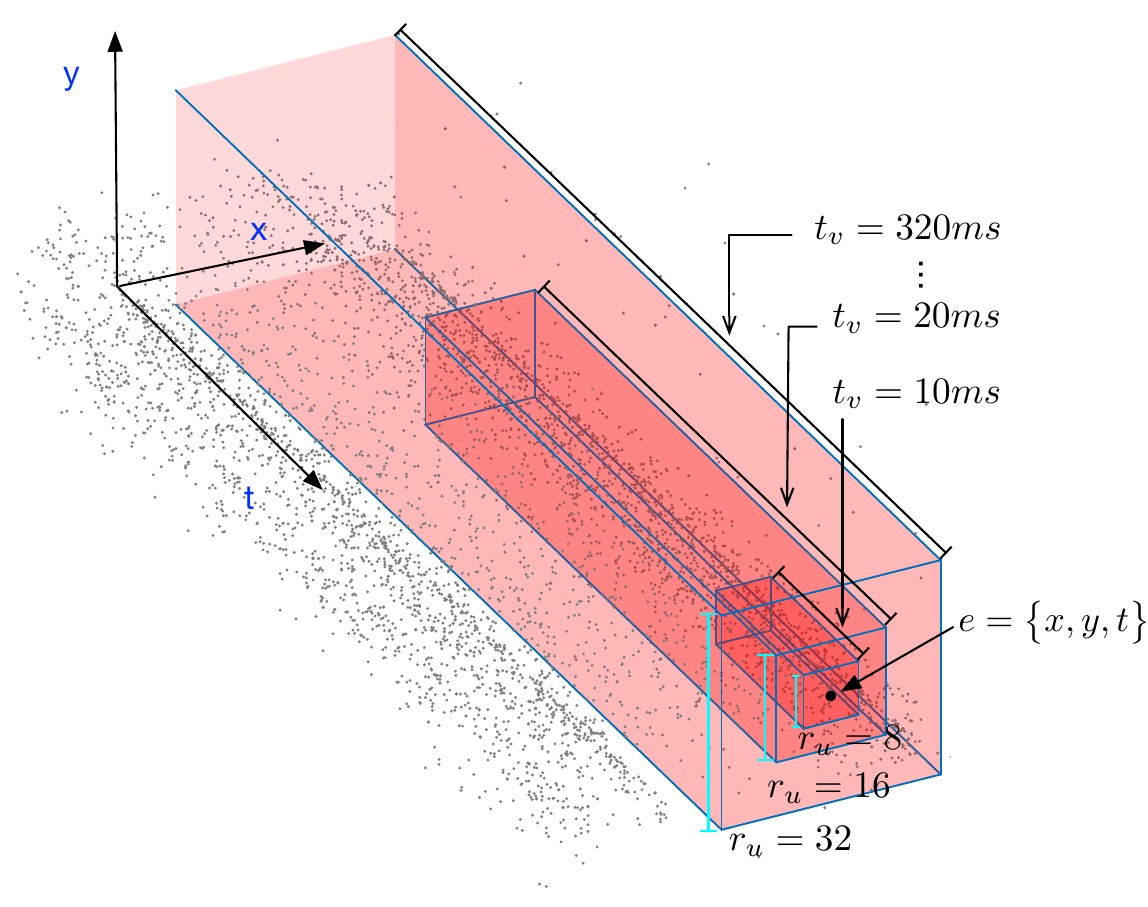}
\caption{
Illustration of the spatio-temporal neighborhoods for an event $e=(x,y,t)$, used to compute the response of the sum of events filters. Past events (black dots) that precede $e$ are summed over time windows of length $t_u \in \{10, 20, 40, 160, 320\}$ms and over spatial windows of radius $r_u \in \{1,2,4,8,16,32\}$ .
\label{fig::salient_events}}
\end{figure}

Saliency values are updated asynchronously at each new event, with all events that occurred in the immediately preceding time windows defined by $t_u$ and within the spatial windows defined by $r_u$ as in eq~\ref{eq:SST} and as illustrated in figure~\ref{fig::salient_events}. We call this the event-based Spatio-Temporal (evST) model.

\section{Performance evaluation tools}
\label{sec:evaluation_tools}
\subsection{Saliency metrics}

To obtain an objective performance metric for computational models of covert selective attention, Parkhurst et al. proposed to evaluate how well a model predicts eye movements, i.e.,  overt attention~\cite{Parkhurst_etal02a}. This has become the standard method in the field which we also adopt. The paradigm allows for multiple metrics for the comparison between model predictions and fixation locations. One class of metrics used for this purpose are \textit{Location-based}; examples are the Area Under the Curve (AUC)~\cite{Green1966,Emami2013,Engelke2013} and its variation. the shuffled AUC (sAUC)~\cite{Borji2013a,
Borji2013c,Li2015a}, the Normalized Scanpath Saliency (NSS)~\cite{Itti2005,Peters2005}, and information gain (IG)~\cite{Kummerer2014,Kummerer2015}. Another class of metrics are \textit{Distribution-based} such as the similarity metric (SIM), known also as \textit{histogram intersection}, the Pearson's Correlation Coefficient (CC)~\cite{Pearson1894}, and the \textit{Kullback-Leibler divergence} (KL-Div)~\cite{LeMeur2007,Wilming2011}.

It has been suggested ~\cite{Riche2013,Bylinskii2019} that a reliable comparison of computational models with human fixations requires to use more than one of these metrics. We therefore evaluate our model with two location based metrics (NSS and sAUC) and two distribution based metrics (SIM and CC).

\subsection{Normalized Scanpath Saliency}
The Normalized Scanpath Saliency metric~\cite{Peters2005} is a discrete  measure of the correlation between fixated locations and their saliency. It is parameter-free and defined by 

\begin{equation}
\text{NSS}=\frac{1}{N}\sum^{N}_{i=1} \frac{S_i-\bar{S}}{\sigma_{s}},
\end{equation} 
where $S_i, i=1,\ldots,N$ are the locations of the $N$ fixations and  $\bar{S}$ and $\sigma_s$ are, respectively, the mean and the standard deviation of the saliency map $S$. For $NSS=1$, the salience at fixation locations is on average one standard deviation above the average while for $NSS=0$, the saliency map prediction is only as good as chance. Note that the smaller the standard deviation of a saliency map, the better the NSS metric predicts the fixations locations. Finally, due to the centering of the NSS by subtracting the mean, the NSS is invariant to linear transformations of the saliency map, e.g., if the map is considered as a distribution, its normalization has little impact on the NSS score~\cite{Bylinskii2019}. 

\subsection{Shuffled Area under the ROC Curve}

Let us define the correct prediction of a fixation by a computational attention model as a hit (true positive) and the incorrect prediction that a non-fixated pixel is a fixation, while in fact it is not, as a false alarm (false positive). Plotting the true positive rate against the false positive rate is called the Receiver Operating Characteristic (ROC). Its integral, or Area Under the Curve (AUC) provides a natural saliency metric that considers the saliency map as a classifier for pixels being fixations or not~\cite{Green1966,Emami2013,Engelke2013}. To account for biases common to all (or many) images, e.g. the tendency of many observers to focus on the center of a scene, we use the shuffled AUC (sAUC). Zhang et al.~\cite{Zhang_etal08b} constructed this metric by defining the hits as defined above and the false alarms as the union of all
fixations of all subjects across all other images, except for the hits in the currently viewed scene.

\subsection{Similarity}
The similarity metric (SIM) was introduced in image content based matching by Swain and Ballard~\cite{Swain1991} to quantify the intersection of histograms. It was widely adopted due to its simplicity:
\begin{equation}
\text{SIM}(S,FM) = \sum_i \min(S_i,FM_i),
\label{eq:SIM}
\end{equation}
where, in our application, $S_i$ and $FM_i$ are respectively the $i$-th pixels of the saliency and the fixation map to be compared, and the sum runs over all pixels of the maps. Both maps are normalized as distributions, i.e. the sums of all their elements are equal to unity.
Perfect agreement of $S$ and $FM$ results in $\text{SIM}(S,FM)=1$ while completely different maps give $\text{SIM}=0$.

\subsection{Pearson's correlation coefficient}
Pearson's correlation coefficient is a well established statistical metric used to quantify how closely two variables are correlated. It is defined as:
 \begin{equation}
CC(S,FM)=\frac{cov(S,FM)}{\sigma(S) \times \sigma(FM)}, \end{equation}
where $cov$ is the covariance and $\sigma$ is the standard deviation. S and FM have the same definitions as in the similarity metric.

\section{Ground-truth data for overt attentional selection}
\subsection{Simultaneous acquisition of event based and frame based dynamic scenes}

Videos of indoors and outdoors scenes were acquired with an ATIS camera which records both change events
and gray level events, defined as a pixel's intensity when it undergoes a change event. To generate the frame based videos shown to the participants, gray level images are read from the gray level events buffer at 100 frames per second, in agreement with the  temporal accuracy of the eye tracker. Simultaneously, for comparison with the frame-based saliency models in section~\ref{sec:Results}, saliency maps are also generated at 100 frames per second from the saliency 2D-buffer. 

\begin{figure}[H]
    \subfloat[Foosball (138)]{\includegraphics[width=0.16\textwidth]{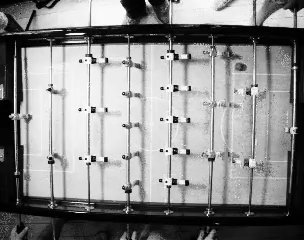}}\hfill
    \subfloat[Bastille (67)]{\includegraphics[width=0.16\textwidth]{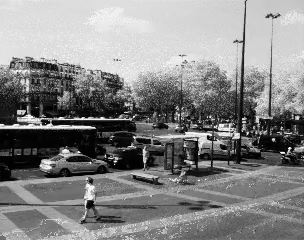}}\hfill
    \subfloat[Objects (128)]{\includegraphics[width=0.16\textwidth]{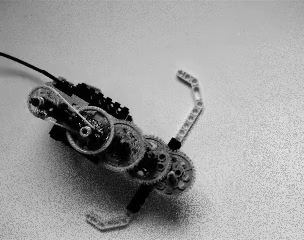}}\hfill
    \subfloat[People (26)]{\includegraphics[width=0.16\textwidth]{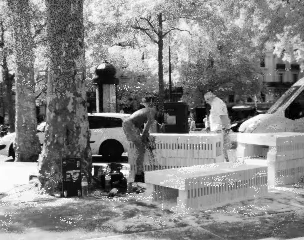}}\hfill
    \subfloat[Shapes (42)]{\includegraphics[width=0.16\textwidth]{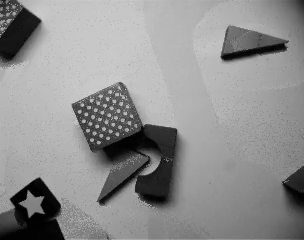}}\hfill 
    \subfloat[Street (197)]{\includegraphics[width=0.16\textwidth]{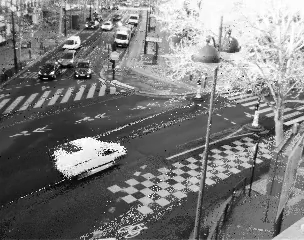}}
    \caption{Sample frames from each type of video category. Numbers in parentheses give the number of videos for each category (total 598).}
    \label{f:samples}
\end{figure}

All videos were recorded with a stationary ATIS camera  observing dynamic scenes of moving physical objects (no simulations). Scenes from six categories were recorded; an example of a still image of a scene from each category is shown in figure~\ref{f:samples}.

\subsection{Eye-tracking}
\label{sec:eyetracking}
The video material was used to produce 598 short videos that were displayed to a group of human observers while their eye movements were tracked.
Each participant watched 300 videos, randomly selected out of the corpus of 598. Twenty participants (10 female, 10 male) with ages between 20 and 54 years (average = 35 years, std dev = 9.98 years) participated in the study which was approved by the Institutional Review Board from the Sorbonne University and executed in the StreetLab~\cite{StreetLab}. Participants had normal or corrected to normal vision and gave written or verbal consent. To limit top-down effects these clips were limited to a duration of eight seconds each.

\begin{figure}
\centering
\subfloat[]{\includegraphics[width=.98\columnwidth]{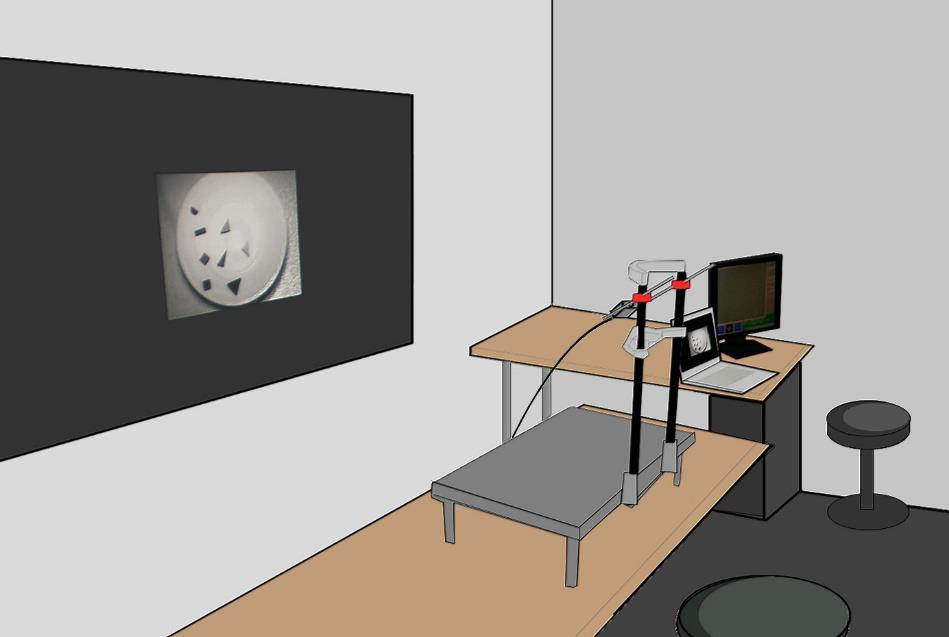}}\hfill
\subfloat[]{\includegraphics[width=.98\columnwidth]{chinrest.png}}\hfill
\caption{Schematic view of eye-tracking acquisition setup. (a) The left stool was for the volunteer while the operator sat on the stool on the right, controlling the experiment. (b) The eye-tracking cameras from the Eyelink II system were fixed to the chin rest holder. 
}
\label{f:streetlab}
\end{figure}

Participants were seated on a height-adjustable stool facing a 168 $\times$ 94 cm screen, with their chin on a chin rest, figure \ref{f:streetlab}. This screen was \SI{1,50}{\meter} from the chin rest and had a \SI{100}{\hertz} refresh rate and a resolution of 1024 $\times$ 768 pixels. Since the ATIS camera has a spatial resolution of 304 $\times$ 240 pixels only the center area of the screen was used, with the surrounding area dark (black). 

Participants were naive to the goals of the study. The experiment took approximately one hour and a half per participant including breaks, an introduction to the free-viewing paradigm and audio recordings (see below). Viewing time was divided in ten seven-minute blocks. Each block started with a 9-point calibration followed by 30 successive alternations of a central fixation point and an eight second video.

Between blocks, participants were encouraged to share three things they saw in the videos, their answers were recorded with a microphone. No other instructions were given to the participants. The purpose of asking these questions was two-fold: to provide a break in the viewing task, and to maintain the participants' attention directed to the videos. The audio recordings could also be used determine any potential correlations between the vocal responses and the fixation patterns. This has not been explored in the current study.

While participants were watching the videos, their eye positions were recorded using a modified Eyelink II (SR Research, Ottawa, Canada) eye-tracking system. The Eyelink headset has three cameras: one pointing towards each eye and one mounted on the forehead to track the global head position. For this experiment we removed the eye-tracking cameras from the head-set and attached them to the chin rest.  The third (forward-facing) camera was not used. 

\section{Results}
\label{sec:Results}
We first quantitatively assess, in Section~\ref{sec:quantitative},  to what extent participants overtly attend the videos. 
In Section \ref{sec:temporalscales}, we explore the influence of different temporal scales over which events are integrated. In Section~\ref{sec:metrics} we use the four metrics introduced above (NSS, sAUC, SIM and CC) to compare the fixation-prediction performance of the event-based Spatio-Temporal (evST) method with two models from the literature. The first is the Itti \textit{et al } saliency map model~\cite{Itti1998}. We used the simplified implementation ("simpsal") from~\cite{Harel2006}\footnote{\url{http://www.animaclock.com/harel/share/gbvs.php}} and augmented it with a motion channel and a flicker channel~\cite{Niebur_Koch96b,Itti2004}.  The second model is a state-of-the art proto-object based saliency model~\cite{Molin_etal21}. By design this model includes motion information through the use of spatiotemporal filters on sequences of frames. We used the code provided by the authors of that study\footnote{\url{https://github.com/csmslab/dynamic-proto-object-saliency}}. Note that because frames are generated by the gray level events from the ATIS, the videos watched by the human participants were gray-level. Therefore, color channels were inactive in all of the models, including the weakly phasic channel in~\cite{Molin_etal21} that is based on color information.

\subsection{Quantitative assessment of attention to stimuli}
\label{sec:quantitative} 

The fixations from the eye-tracker provide the ground truth against which we compare three saliency models. 
We first provide an analysis of the robustness of these ground truth data.

\subsubsection{Visual attention} 
\begin{figure}
\includegraphics[width=\columnwidth]{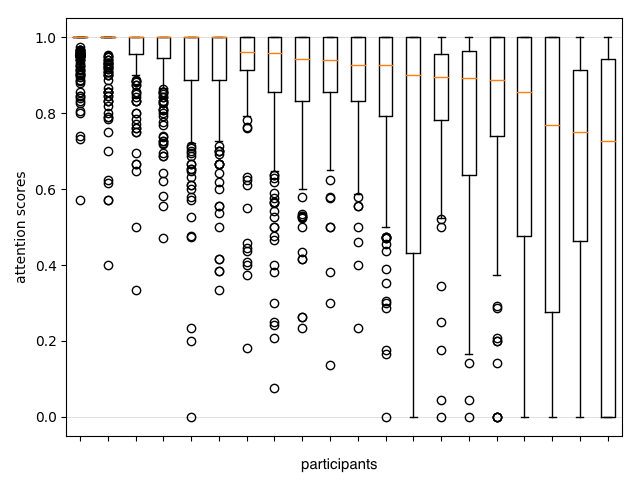}
\caption{Box plot of the variability in attention scores per video for each participant.
Participants are ordered by decreasing mean attention score 
(red lines).}
\label{f:attention_box}
\end{figure}

Fixations were on average \SI{413}{\milli\second} long (std dev = \SI{458}{\milli\second}). Some were not on the video but on the black area of the screen surrounding the video display, see Section~\ref{sec:eyetracking}.
We consider these "inattentive" fixations and use them to compute a visual attention score and determine whether a participant's data should be included in the analysis. We define an attention score as the ratio of attentive fixations over total fixations.
Figure \ref{f:attention_box} shows great variability in the fixation scores between participants. However, even for the least attentive participant there are videos for which all fixations are in the region of interest. Videos with low attention scores were removed from the analysis (see below), but this did not result in the removal of all data from any of the participants. 
Even for the "least attentive" participant, on average the majority of fixations were on the video portion of the screen.

\begin{figure}
\includegraphics[width=\columnwidth]{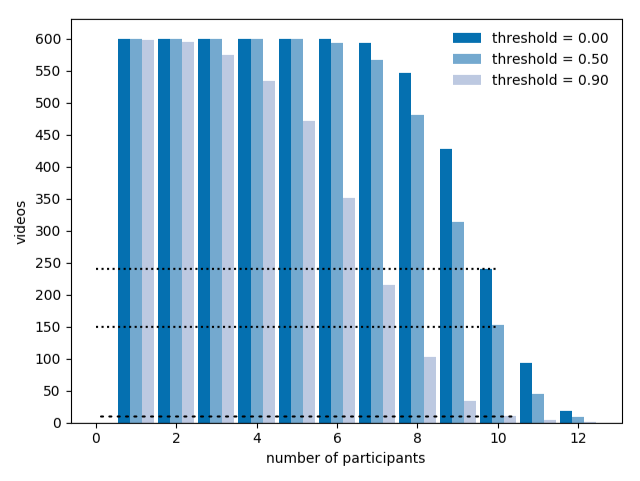}
\caption{
Cumulative histogram showing the minimum number of participants surpassing the plotted attention scores.  At least 10 participants saw 240 videos (upper dashed line), 152 of those were video clips with an attention score exceeding a threshold of 0.5 (middle dashed line) and 11 with an attention score of 0.9 (lower dashed line.
}
\label{f:attention_hist}
\end{figure}

We decided to eliminate data from videos with low attention scores. We must choose a compromise between a small but high quality dataset (given by a high threshold) and a larger but lower-quality dataset (low threshold). This is illustrated in Figure~\ref{f:attention_hist}: For instance, there are 240 videos that are seen by 10 or more participants
(upper dashed line). If we remove those with an attention score lower than 0.5,  152 remain (middle dashed line); if we remove all videos with an attention score lower than 0.9, only 11 remain (lower dashed line). We opt for a high-quality data set with an attention threshold of 0.9, all following results are obtained from this data set. It comprises a total of 57,798 fixations from 20 participants in 598 videos.

There are a few other informal observations, not discussed in detail here. First, the experiment took less time for the participants who focused on the videos. For those whose gaze drifted, more re-calibrations were necessary, increasing the duration of the experiment. We also noticed that the frequency of inattentive fixations increased over the course of the experiment, presumably due to fatigue.

\subsection{Integration over time}
\label{sec:temporalscales}

The impact of the time window on the saliency metrics is expected to be significant as events are triggered by scene contents. Scene dynamics, i.e. its changes and rates of changes, can be correctly captured if the adequate time window is used. This section evaluates the behavior of the saliency metrics for increasing temporal window size $t_u$ (eq~\ref{eq:scales}) while summing over all spatial scales. Therefore,  for each value of $t_u$ we compute a map for $\displaystyle \sum_{v=0}^5 S_{u,v}$ and normalize it to the range $[0,1[$ as before.\\

As mentioned in section~\ref{sec:model}, the saliency computation pipeline uses events stored in a 2D limited memory buffer that inherently provides some temporal 
adaptation in the computation of the spatio-temporal saliency. By changing manually $t_u$, we can more explicitly examine the temporal contribution of events by excluding the ones not consistent (e.g. too old) with the stimulus dynamics. 

Due to the large amount of saliency maps to be generated, we conducted this test only over a subset of each category of the database. We randomly selected 10\% of videos from each category for each of the six values of $t_u$, increasing from 10ms to 320ms and for the three models.

Table~\ref{Tab:MetricTempScale} summarizes the results on the four metrics and also the results achieved by the full form of the evST model that integrates over all the time windows. It shows that three of the four metrics are increasing functions of $t_u$, the exception being sAUC where in several scene categories the model performances was best for some shorter time scales than for the longest one. However, for this and indeed for all four metrics, the complete evST model (with integration over all time scales) tends to either achieve the highest scores or is on par with best score from fixed time windows.  As the full evST model provides in most cases the best performance, we will use the full model. This is not adding much computational complexity when compared to the use of only one time window. 

\begin{table*}[ht]
\begin{tabular}{p{.5\textwidth} p{.5\textwidth}}
\multicolumn{1}{c}{NSS} & \multicolumn{1}{c}{sAUC}\\[0.25ex]
\resizebox{.5\textwidth}{!}{
\begin{minipage}{.525\textwidth}
\sisetup{round-mode=places,round-precision = 3, detect-weight, mode=text, table-format=+1.3} 
\centering
\begin{tabular}{c S[table-format=1.3] S S[table-format=1.3]S[table-format=1.3] SS}  
\toprule
 $t_u$ (ms) & {Foosball} & {Bastille} & {Objects} & {People} & {Shapes} & {Street} \\
\midrule
10 & 0.7246& 0.4558 & 0.6648 & 0.5439 &0.2602 & 0.3936\\
20 & 0.7806& 0.4662 & 0.7481 & 0.6151 &0.3715 &0.4014\\
40 & 0.8175& 0.4752 & 0.8307 & 0.6862 &0.4507 &0.4341\\
80 & 0.8728& 0.4904 & 0.8996 & 0.6828 & 0.5345 & 0.4579\\
160& 0.9138& 0.5072 & 0.9504 & 0.7550 & 0.6607 & 0.4957\\
320& \boxed{0.952}& \boxed{0.509} & \boxed{0.998} & \boxed{0.775} & \boxed{\bf 0.810} & \boxed{\bf 0.534}\\
\midrule
evST & \bf 1.0108& \bf 0.5550& \bf 1.0597& \bf 0.8316&  0.5946 &  0.4790\\
\bottomrule
\end{tabular}
\end{minipage}} 
&
\resizebox{.47\textwidth}{!}{
\begin{minipage}{.525\textwidth}
\sisetup{round-mode=places,round-precision = 3, detect-weight, mode=text, table-format=+1.3} 
\centering
\begin{tabular}{c S[table-format=1.3] S S[table-format=1.3]S[table-format=1.3] SS}  
\toprule
 $t_u$ (ms) & {Foosball} & {Bastille} & {Objects} & {People} & {Shapes} & {Street} \\
\midrule
10 & 0.6881& 0.6032& 0.6157& 0.6102& 0.5755 &0.5514\\
20 & 0.6645& 0.5955& 0.6122& 0.6191& 0.6098 &0.5532\\
40 &  \boxed{0.689} & 0.6064 & \boxed{0.619} & 0.6112& 0.5930& 0.5461\\
80 & 0.6592 & 0.5920 & 0.5986 & \boxed{0.630} & 0.6239 &0.5617\\
160 & 0.6849 & 0.6035 & 0.6100 & 0.6017& 0.6332 & \boxed{\bf 0.564}\\
320 & 0.6818 & \boxed{0.608} & 0.6125 & 0.6102& \boxed{0.640} &0.5583\\
\midrule
evST & \bf 0.7154& 0.60250 & \bf 0.6198 & \bf 0.6460 & \bf 0.6610& \bf 0.5635\\
\bottomrule
\end{tabular}
\end{minipage}}\\[1ex]
 & \\
\multicolumn{1}{c}{SIM} & \multicolumn{1}{c}{CC}\\[0.25ex]
\resizebox{.5\textwidth}{!}{
\begin{minipage}{.525\textwidth}
\sisetup{round-mode=places,round-precision = 3, detect-weight, mode=text, table-format=+1.3} 
\centering
\begin{tabular}{c S[table-format=1.3] S S[table-format=1.3]S[table-format=1.3] SS}  
\toprule
 $t_u$ (ms) & {Foosball} & {Bastille} & {Objects} & {People} & {Shapes} & {Street} \\
\midrule
10  & 0.3579 & 0.3861 & 0.3422 & 0.2686 & 0.1929 &0.2587\\
20  & 0.3809 & 0.4023 & 0.3769 & 0.3205 & 0.2419 &0.3039\\
40  & 0.4007 & 0.4157 & 0.4059 & 0.3581 & 0.2887 &0.3395\\
80  & 0.4184 & 0.4293 & 0.4307 & 0.3888 & 0.3321 &0.3696\\
160 & 0.4341 & 0.4435 & 0.4466 & 0.4052 & 0.3697 &0.3948\\
320 & \boxed{\bf 0.448} & \boxed{\bf 0.457} & \boxed{0.459} & \boxed{0.417} & \boxed{\bf 0.400} &\boxed{\bf 0.422}\\
\midrule
evST & 0.3979& 0.4362 & \bf 0.4935 & \bf 0.4328 & 0.3853 &0.3999\\
\bottomrule
\end{tabular}
\end{minipage}}
&
\resizebox{.47\textwidth}{!}{
\begin{minipage}{.525\textwidth}
\sisetup{round-mode=places,round-precision = 3, detect-weight, mode=text, table-format=+1.3} 
\centering
\begin{tabular}{c S[table-format=1.3] S S[table-format=1.3]S[table-format=1.3] SS}  
\toprule
 $t_u$ (ms) & {Foosball} & {Bastille} & {Objects} & {People} & {Shapes} & {Street} \\
\midrule
10  & 0.1954 & 0.1743 & 0.2964 & 0.2080 & 0.0562 &0.1446\\
20  & 0.2158 & 0.1847 & 0.3343 & 0.2365 & 0.0722 &0.1680\\
40  & 0.2377 & 0.1927 & 0.3731 & 0.2547 & 0.0885 &0.1875\\
80  & 0.2569 & 0.2008 & 0.4092 & 0.2707 & 0.1057 &0.2064\\
160 & 0.2739 & 0.2106 & 0.4359 & 0.2841 & 0.1218 &0.2220\\
320 & \boxed{\bf 0.294} & \boxed{\bf 0.219} & \boxed{0.457} & \boxed{0.294} & \boxed{\bf 0.138} &\boxed{0.243}\\
\midrule
evST& 0.2704 & 0.2167 & \bf 0.4904 & \bf 0.3134 & 0.1221 &\bf 0.2503\\ 
\bottomrule
\end{tabular}
\end{minipage}}\\[1ex]
 & \\
\end{tabular}
\caption{Impact on the four metrics of the temporal windows for $t_u\in [10,320]$ ms, tested over subsets of each categories. Larger is better for all metrics.
Best scores with a fixed time window are highlighted with boxes and best scores for each metric are in bold font. 
\label{Tab:MetricTempScale}}
\end{table*}


\subsection{Performance comparisons between models}
\label{sec:metrics}
The four metrics are shown for each video and for each saliency model in figures~\ref{f:nss_box_vid} to~\ref{f:CC_all}. Table~\ref{Tab:MetricAllComp} provides an assessment of the average performance on each video category and each metric.

\subsubsection{NSS}
\begin{figure*}
\includegraphics[width=\textwidth]{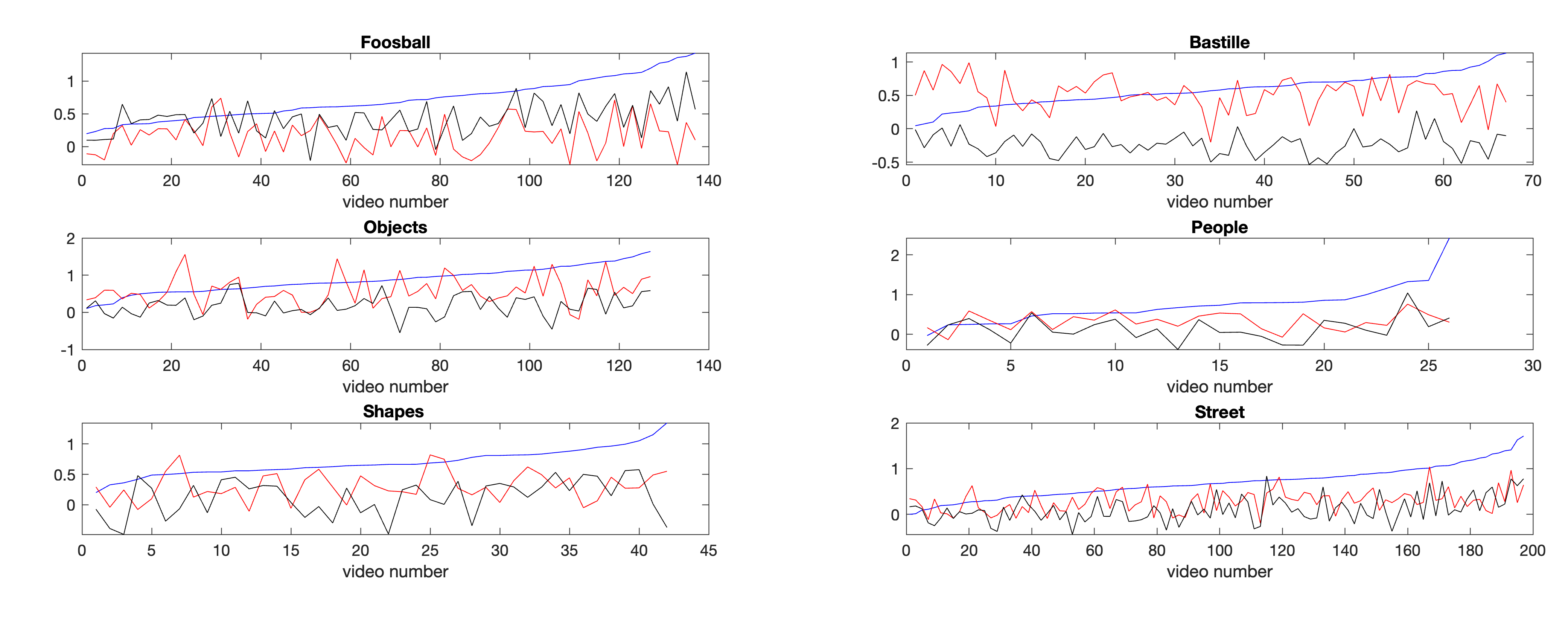}
\caption{NSS scores for all videos for all three models and each category. Scores are plotted in ascending order of the evST model (blue). In red and black are shown results from the Proto Object model and of the Itti  et al model, respectively. An NSS score of zero corresponds to random choices.
\label{f:nss_box_vid}}
\end{figure*}

The NSS score, for almost all videos, is highest for the evST saliency model, as shown in figure~\ref{f:nss_box_vid} and in the averaged scores in Table~\ref{Tab:MetricAllComp}. There is only one video for which the evST model gives a negative score. In contrast, the Proto object and the Itti et al models have, respectively, 96 and 243 videos with negative scores. The evST model fares better in \SI{82}{\percent} of the videos  than the other two models. 

The closest competition to the evST model is from the Proto-object model in the "Bastille" sequence. Here, there are several videos where the performance of the Proto-object model clearly exceeds that of the evST model. However, even in this video category, the evST model shows overall clearly superior performance, as is seen in Table~\ref{Tab:MetricAllComp}.

\subsubsection{sAUC}
\begin{figure*}
\includegraphics[width=\textwidth] {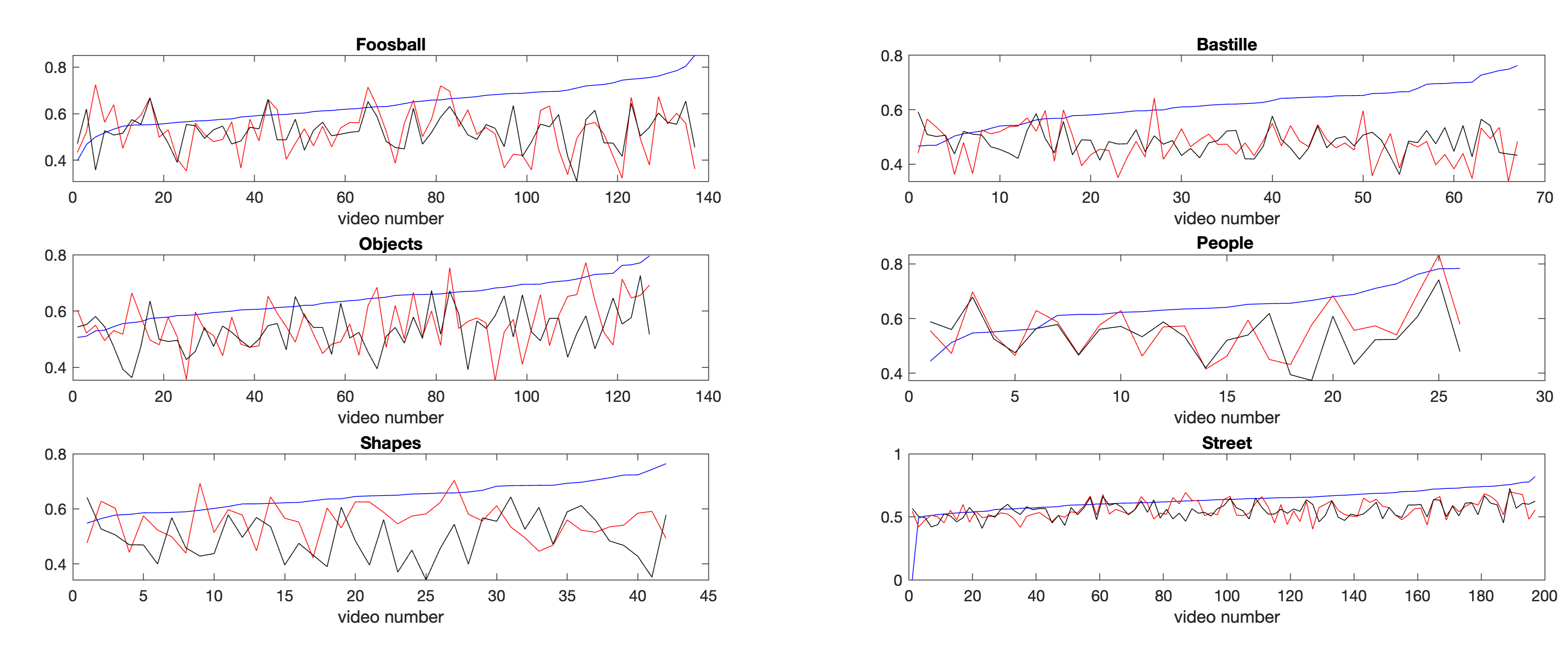}
\caption{sAUC scores for all videos for all three models. Notations are as in Fig~\ref{f:nss_box_vid}. An sAUC value of 0.5 means the saliency model is close to random guesses. 
\label{f:sAUC_all}}
\end{figure*}

For this metric as well, the evST model achieved a better performance on average than the two other models as shown in figure~\ref{f:sAUC_all} and Table~\ref{Tab:MetricAllComp}. The evST model performs better than the Proto object model and the Itti et al. model for \SI{92}{\percent} of the videos. 

\subsubsection{SIM}
\begin{figure*}
\includegraphics[width=\textwidth] {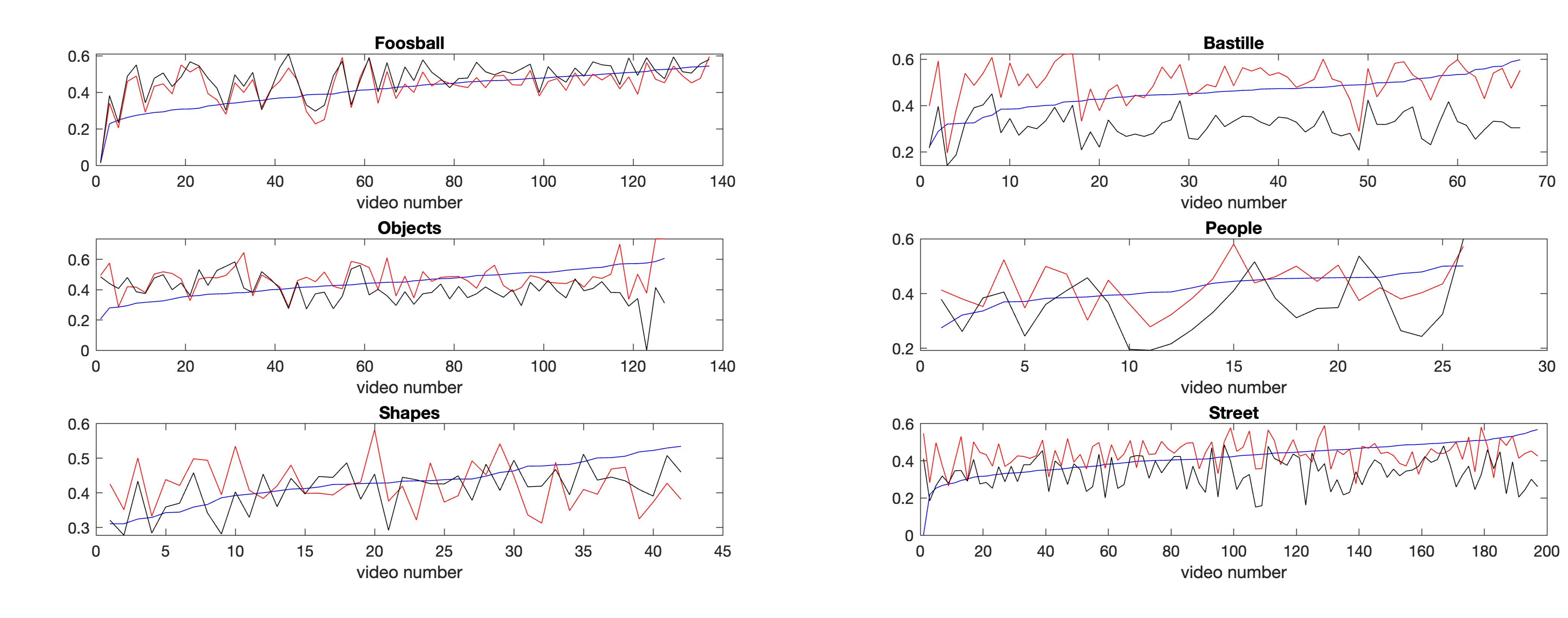}
\caption{SIM scores for all videos given for all three models. Notations as in Fig~\ref{f:nss_box_vid}.
\label{f:SIM_all}}
\end{figure*}

This metric --and the CC metric as well-- does not directly evaluate saliency value at the pixel level, but instead compares globally the saliency maps $S$ against the distribution of fixations $FM$ for each video and for all participants. By this metric, the Proto object model is superior for all sequences but "Foosball", where the Itti et al. model achieves best, as shown by figure~\ref{f:SIM_all} and Table~\ref{Tab:MetricAllComp}. The evST model does not achieve the highest average score in any of the video categories.

\subsubsection{CC}
\begin{figure*}
\includegraphics[width=\textwidth] {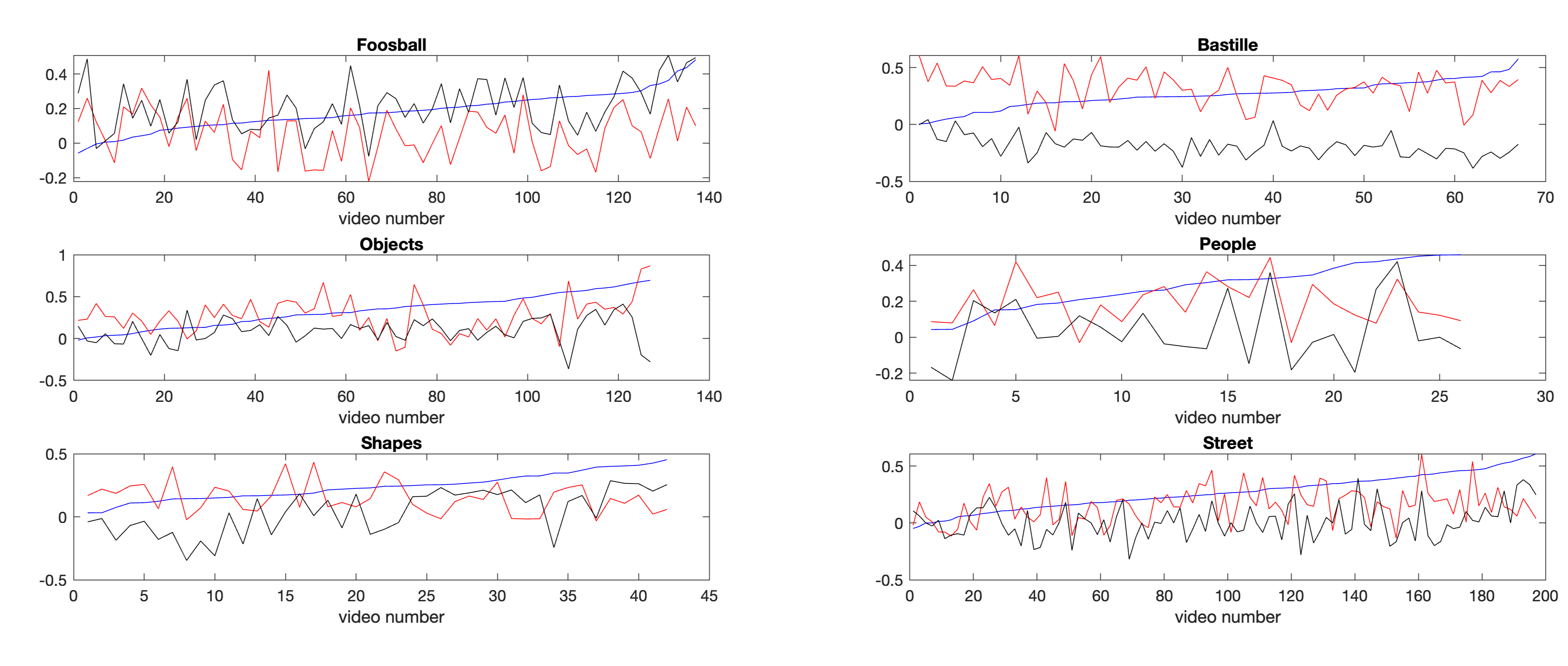}
\caption{CC scores for all videos for all three models. Notations as in Fig~\ref{f:nss_box_vid}. Zero value means chance performance.}
\label{f:CC_all}
\end{figure*}

Each of the three models outperforms the others in at least one video category by this metric, as shown by figure~\ref{f:CC_all} and Table~\ref{Tab:MetricAllComp}. However, the evST model overall dominates the results, with highest scores in the majority (4/6) of categories while the other two models  are best in only a single category each.

\begin{table*}
\begin{tabular}{p{.48\textwidth} p{.48\textwidth}}
 \multicolumn{1}{c}{NSS} & \multicolumn{1}{c}{sAUC}\\[0.25ex]
\resizebox{.5\textwidth}{!}{
\begin{minipage}{.55\textwidth}
\sisetup{round-mode=places,round-precision = 3, detect-weight, mode=text, table-format=+1.3} 
\begin{tabular}{c S[table-format=1.3] S S[table-format=1.3]S[table-format=1.3] SS}  
\toprule
  & {Foosball} & {Bastille} & {Objects} & {People} & {Shapes} & {Street} \\
\midrule
evST & \bf 0.721& \bf 0.575& \bf 0.873 &\bf 0.731& \bf 0.694& \bf 0.689\\
Proto object& 0.1639& 0.5154& 0.5612& 0.3244& 0.3060& 0.3111\\
Itti& 0.4231& -0.2325& 0.2031& 0.1302& 0.1271& 0.1151\\
\bottomrule
\end{tabular}
\end{minipage}}
& 
\resizebox{.5\textwidth}{!}{
\begin{minipage}{.55\textwidth}
\sisetup{round-mode=places,round-precision = 3, detect-weight, mode=text, table-format=+1.3} 
\begin{tabular}[t]{c S[table-format=1.3] S S[table-format=1.3]S[table-format=1.3] SS}
\toprule 
 & {Foosball} & {Bastille} & {Objects} & {People} & {Shapes} & {Street} \\
\midrule
evST & \bf 0.641& \bf 0.615& \bf 0.643 &\bf 0.637&\bf 0.647& \bf 0.635\\
Proto object & 0.5206& 0.4755& 0.5588& 0.5621& 0.5533&0.5491\\
Itti & 0.5285& 0.4821& 0.5224& 0.5387& 0.4960& 0.5417\\
\bottomrule
\end{tabular} 
\end{minipage}} \\[1ex]
 & \\
\multicolumn{1}{c}{SIM} & \multicolumn{1}{c}{CC}\\[0.25ex]
\resizebox{.5\textwidth}{!}{
\begin{minipage}{.55\textwidth}
\sisetup{round-mode=places,round-precision = 3, detect-weight, mode=text, table-format=+1.3} 
\begin{tabular}{c S[table-format=1.3] S S[table-format=1.3]S[table-format=1.3] SS}  
\toprule
  & {Foosball} & {Bastille} & {Objects} & {People} & {Shapes} & {Street} \\
\midrule
evST & 0.4010 & 0.4516 & 0.4351 &.4221 &0.4175 & 0.4056\\
Proto object & 0.4225 & \bf 0.502 & \bf 0.479& \bf 0.425& \bf 0.423& \bf 0.440\\
Itti &\bf 0.466& 0.3144& 0.4051& 0.3541& 0.4111& 0.3327\\\bottomrule
\end{tabular}
\end{minipage}}

& \resizebox{.5\textwidth}{!}{
\begin{minipage}{.55\textwidth}
\sisetup{round-mode=places,round-precision = 3, detect-weight, mode=text, table-format=2.3} 

\begin{tabular}[t]{c S[table-format=1.3] S S[table-format=1.3]S[table-format=1.3] SS}
\toprule 
 & {Foosball} & {Bastille} & {Objects} & {People} & {Shapes} & {Street} \\
\midrule
evST & 0.1833 & 0.2617 & \bf 0.332 & \bf 0.281 & \bf 0.238 & \bf 0.269\\
Proto object &  0.0804 & \bf 0.324 & 0.2871& 0.1891& 0.1478& 0.1505\\
Itti & \bf 0.232& -0.1896& 0.0958& 0.0388& 0.0382& 0.0017\\
\bottomrule
\end{tabular} 
\end{minipage}}\\[1ex]
 & \\
\end{tabular}
\caption{All four metrics averaged scores per video for the three models. Larger is better and highest scores are in bold.
\label{Tab:MetricAllComp}}
\end{table*}

\section{Discussion and Conclusion}
We compare the predictive power of our newly developed saliency model with that of two established computational models. We selected these models because they were developed for explaining neuronal mechanisms of visual selective attention, rather than being fine-tuned (trained) to perform optimally in eye-movement prediction benchmark tests. In fact, the two models we compare evST with were not trained at all, and obviously no training is involved in the evST model itself.

We find that evST is highly competitive with both of these models. As Table~\ref{Tab:MetricAllComp} shows, evST shows the highest performance in all scene categories for two of the metrics we used (NSS and sAUC) and in the majority (four out of six)
of scene categories in a third one (CC). In the fourth category (SIM), the Proto-object model dominates the other two but evST is second-best in four of the six scene categories. Over all scene categories and metrics, evST obtained the highest score in 16 out of 24 categories, far surpassing both other models (Proto-object:~6, Itti et al:~2). 

We acknowledge that we had to use  slightly simplified versions of the two comparison models because both use color contrast as one of their features. Since our recordings are gray-scale, we simplified both models by removing the corresponding color-based feature maps. We also point out that the original Itti et al model~\cite{Itti1998} operates entirely on static scenes, i.e., it considers image frames as independent and ignores all temporal relationships between them, including differences between frames. For this reason, we felt it would be "unfair" (favoring the evST model) to compare the evST model with the original model using dynamic scenes. Therefore, we used an extended version of the Itti et al model which is augmented by both a flicker and a motion component. Nevertheless, even with these augmentations, the evST model predicts eye fixations substantially better than the Itti et al model. More surprising that this "classic" model is surpassed in performance by the evST model is that the latter overall also exceeds the performance of a state-of-the-art  proto-object based model~\cite{Molin_etal21}, though by a smaller margin. In addition to taking into account Gestalt-laws of perceptual organization to structure the static feature maps of earlier models, this model also includes dynamic (motion) features. It does not, however, include localized temporal differences (flicker), the \textit{only} feature present in the evST model. This feature was introduced in a very early saliency map~\cite{Niebur_Koch96b,Niebur_Koch95b} but, perhaps surprisingly, omitted in many later models.

While our benchmarking results indicate a highly credible performance of evST for predicting human saccades, as mentioned previously our emphasis is not on optimizing benchmark numbers. Instead, our goal is to develop and understand the mechanisms that control the deployment of visual selective attention, specifically bottom-up attention that is determined by saliency. We feel that there are two reasons why the event-based signal is highly promising for use as a saliency indicator. 

The first is  efficiency. Compared to other feature maps, computation of local change (flicker) requires few computational resources even in standard frame-based technology (e.g., different from many commonly used feature types contributing to saliency, no convolutions are required to identify areas with temporal change, only a pixel-wise difference between frames). We do not generate frames of saliency maps, in fact the notion of an image "frame" has as little meaning in our model as it has in biology.  Forgoing all the related complexities and constraints associated with standard image frame-based technology  we can update the saliency of each pixel with microsecond precision.

This advantage is highly amplified in event-based neuromorphic technology. Our saliency measure is obtained with minimal computational effort, essentially counting local events in the immediate event history followed by a simple  normalization step, eq~\ref{eq:SST}. We thus propose that the raw event signal originating in event-based vision sensors followed by minimal computation can serve as a first approximation of a saliency map. Saliency is thus obtained nearly "for free" (for a related idea in the domain of saliency in biological vision see~\cite{VanRullen2003}).

What is the price to pay for this nearly unsurpassable level of efficiency? The most obvious is that temporal change is only one component out of several that make a visual scene element salient;  other submodalities clearly contribute to saliency. This is most obvious in still images where using  time-invariant visual features exclusively is sufficient for computational models of selective attention to make predictions that are  significantly better than chance~\cite{Parkhurst_etal02a}. However, in most naturally occurring situations complete lack of motion/change is rare. Itti~\cite{Itti05} found that temporal change (and motion) are more reliable predictors of human saccades than static visual features. While best prediction performance is achieved when all features (static and dynamic) are taken into account, and we therefore pay a price in \textit{efficacy} in the evST model by using only temporal change, this price has to be weighed against the \textit{efficiency} of using the raw event signal directly, with only 
minimal  computational effort and memory load, as discussed above. 

The second, and likely related reason for focusing on temporal change for saliency computation is its high ecological relevance. The first explicit computational saliency map models~\cite{Itti1998,Itti_Koch01} were developed for still images and based on the differences of static features (intensity, color, depth etc) \textit{in space}. In contrast, saliency in the current study is computed entirely from local contrast\textit{ in time}. Both temporal and spatial contrast contribute to computational saliency, with the former having more influence than the latter~\cite{Itti05}. In many areas, responses to transient stimuli are of great importance in biological vision and in neuronal processing in general where phasic (transient) responses are very common. 

In the more narrow context of selective attention, which is the focus of our research, it has been known for a long time that rapid changes in the visual field attract attention. For instance, abrupt visual onsets are known to "capture" attention~\cite{Yantis_Jonides84,Jonides_Yantis88,Schreij_etal08,Ruthruff_etal21}, resulting in preferential processing of visual input at their location, although this capture can be overridden in some cases~\cite{Bacon_Egeth94}.
This is closely related to the phenomenon of exogenous attention. While endogenous attention is guided by  symbolic cues, e.g. an arrowhead pointiing towards the to-be-attended region, exogenous attention is typically controlled by a peripheral cue e.g. a dot or a small bar that is briefly presented next to the target immediately preceding target onset. This results in an involuntary, automatic, fast ($\approx$hundred ms vs several hundreds of ms for endogenous attention) direction of attention to the target area\cite{Anton-Erxleben_Carrasco13}. 

We thus propose that there may be a strong connection between the events we use in the evST model and the behavioral (and neuronal) effects of exogenous attention. This may explain the good performance of a model that is exclusively based on a very restricted amount of information, the raw events indicating the locations where change occurs in a scene. While there are many important aspects of the control of visual selective attention (endogenous attention, mentioned above, or any of the purely spatial features that have dominated early saliency maps) which have not implemented localized temporal change, our results indicate that this extremely simple mechanism may capture a surprisingly large part of a complex function of perception and cognitive. While motivated in biology, we postulate that these mechanisms are equally applicable to the implementation of machine intelligence.\\

{\bf Acknowledgments} Work supported by ONR grant N00014-22-1-2699 and NSF grant 2223725; by French National Research Agency (ANR) grant ANR-20-CE23-0021.

\bibliographystyle{IEEEtran}
\bibliography{references,niebase}

\end{document}

%% file: abbrevs.tex
\long\def\comment#1{} 
\usepackage{xspace}

\newcommand{\eg}[0]{{\em e.g.}\xspace}
\newcommand{\etc}[0]{{\em etc.}\xspace}
\newcommand{\vs}[0]{{\em vs.}\xspace}